\documentclass [12pt]{article}%

\usepackage{amssymb,amsmath}
\usepackage{color}
\usepackage{hyperref}
\usepackage{gastex}

\newtheorem{thm}{Theorem}
\allowdisplaybreaks[1]
\newcommand{\beq}[1]{  \\ {\tiny ({#1})}   \begin{equation} \label{#1} }
\newcommand{\beqa}[1]{ \\ {\tiny ({#1})}   \begin{eqnarray} \label{#1} }
\renewcommand{\beq}[1]{  \begin{equation} \label{#1} }    	
\renewcommand{\beqa}[1]{\begin{eqnarray} \label{#1} }			

\newcommand{\eeq}{\end{equation}}				\newcommand{\eeqa}{\end{eqnarray}}

\renewcommand{\appendix}{\setcounter{section}{0}\renewcommand{\thesection}{\Alph{section}}  			\section*{Appendix} 
}

\newcommand{\rf}[1]{(\ref{#1})}

\def\bd#1{\mbox{\boldmath$\displaystyle\mathbf{#1}$} }
\def\tens#1{\mathbb{\,#1}}											

\def\tr{\operatorname{tr}} 
\def\diag{\operatorname{diag}}

\def\det{\operatorname{det}}
\def\Sym{\operatorname{Sym}}
\def\dim{\operatorname{dim}}
\def\sgn{\operatorname{sgn}}

\def\singlespacing{\baselineskip=13pt}

\begin{document} 

\pagestyle{myheadings}\markright{{\sc Pure shear axes}  ~~~~~~\today}
\singlespacing

\title{Pure shear axes and  elastic strain energy}
\author{Andrew N. Norris\footnote{Rutgers University, Department of Mechanical and Aerospace Engineering, 98 Brett Road, Piscataway, NJ  08854-8058, norris@rutgers.edu}}
\maketitle

\begin{abstract}
It is well known that a state of pure shear has distinct sets of basis vectors or coordinate systems: the principal axes, in which the stress ${\bd \sigma}$ is diagonal, and {\em pure shear bases}, in which $\diag {\bd \sigma} = 0$. The latter is commonly taken as the definition of pure shear, although  a state of pure shear is more generally defined  by $\tr {\bd \sigma} = 0$.  New results are presented that  characterize all possible pure shear bases.  A pair of vector functions are derived  which generate a set of pure shear   basis vectors from any one member of the triad.  The vector functions follow from  compatibility condition for the pure shear basis vectors, and are independent of the principal stress values. 
The complementary types of vector basis have implications for the strain energy of linearly elastic solids with cubic material symmetry: for a given state of stress or strain, the strain energy achieves its extreme values when the material cube axes are aligned with principal axes of stress or with a pure shear basis.   This implies that the optimal orientation for a given state of stress  is with one or the other vector basis, depending as the stress is to be minimized or maximized, which involves the sign of one material parameter. 

\end{abstract}
%

\section{Introduction}

A state  of pure shear is characterized by 
\beq{81}
\tr {\bd \sigma} = 0, 
\eeq
where ${\bd \sigma}$ is the (symmetric) stress tensor. The principal axis form of stress is    
\beq{31}
{\bd \sigma}=\sigma_1\, {\bd e}_1\otimes {\bd e}_1
+\sigma_2\, {\bd e}_2\otimes {\bd e}_2
+\sigma_3\, {\bd e}_3\otimes {\bd e}_3,  
\eeq
where the orthonormal triad $\{{\bd e}_1,{\bd e}_2,{\bd e}_3\}$ is unique up to reflections, i.e. 
 $\{\pm{\bd e}_1,{\bd e}_2,{\bd e}_3\}$, etc. In a state of pure shear the three principal stresses satisfy
\beq{82}
\sigma_1 + \sigma_2+\sigma_3 = 0. 
\eeq
The condition \rf{81} is unaltered under a change of basis, say from 
$\{{\bd e}_1,{\bd e}_2,{\bd e}_3\}$ to 
 $\{{\bd a},{\bd b},{\bd c}\}$.  We define a {\em pure shear basis} as one for which  the diagonal elements of stress vanish, 
\beq{01}
\sigma_{aa}\equiv 
{\bd a}\cdot({\bd \sigma}{\bd a}) = 0,
\qquad
\sigma_{bb}\equiv 
{\bd b}\cdot({\bd \sigma}{\bd b}) = 0,
\qquad
\sigma_{cc}\equiv 
{\bd c}\cdot({\bd \sigma}{\bd c}) = 0. 
\eeq

Coordinate systems with $\diag {\bd \sigma}=0$ have been the subject of recent interest \cite{Belik98,bh04,Ting06a}, focusing on the relation between eqs. \rf{81} and \rf{01},  and on the characterization of all possible  
 pure shear bases.   The three conditions \rf{01} are the conventional definition of pure shear, although \rf{81} is a more general starting point.  The equivalence is well known, and is mentioned, for instance in Love's book (Section 16 of \cite{Love}).  Perhaps the first proof  that eq. \rf{81} implies eq. \rf{01}  was by Gurtin \cite{Gurtin72}, who in turn ascribed the method to Lew. 
B\v{e}lik and  Fosdick \cite{Belik98} offered two distinct proofs,  geometric and algebraic.  They showed that there is a one parameter set of pure shear coordinate systems.   
Boulanger and Hayes \cite{bh04} subsequently gave an interpretation in terms of elliptical sections of the stress ellipsoid. 
Ting \cite{Ting06a} provided a characterization of directions $\bd n$ such that 
$\sigma_{nn}  = 0$ in terms of the total shear  in the plane normal to $\bd n$, $\tau = |{\bd \sigma}{\bd n}|$.

The first objective here is to provide a simple means to characterize all possible pure shear bases. 
What distinguishes the present discussion from the recent notes \cite{Belik98,bh04,Ting06a} is that the pure shear basis vectors are determined in a manner that is, as far as possible, independent of  the values of the principal stresses, $\sigma_1$, $\sigma_2$ and $\sigma_3$.  We show that one member of the orthonormal triad $\{{\bd a},{\bd b},{\bd c}\}$ determines the other two through a cyclic vector function defined relative to the principal stress axes.  
This representation is achieved by considering  the conditions necessary for a given orthonormal triad to be a pure shear triad.  The conditions are independent of the principal stresses, and the vector triad is completely defined in terms of one element, $\bd a$, say. The stress then enters through the condition that  $\sigma_{aa} $ vanish,  which is satisfied if $\bd a$ lies on the intersection of an elliptical cone and  the unit sphere. In this formulation no assumptions, such as the sign or ordering of the principal stresses, are required.   The complete set of pure shear bases follows simply from the vector functions which define the triad.   

The principal axis and pure shear coordinate systems for a state of pure shear have direct implications for strain energy in materials with cubic symmetry.  Thus, the extreme values of the strain energy correspond to aligning the cube axes  with the two distinct sets of orthonormal base vectors: the principal stress axes, and the pure shear axes.  This provides a simpler method than those discussed in \cite{Norris05} for characterizing the orientations of the cube axes which maximize or minimize strain energy for a given state of stress. 

We begin in Section \ref{sec2} with the general result for the pure shear orthonormal bases, which is discussed subsequently. 
 Implications for strain energy are explored in Section \ref{sec3}. 

\section{Pure shear axes}\label{sec2}

\subsection{Generic functions for the basis vectors}
Consider  basis vectors  $\{{\bd a},{\bd b},{\bd c}\}$ such that \rf{01} is satisfied, i.e.  
\begin{subequations}\label{131}
\beqa{131a}
\sigma_1 a_1^2 +\sigma_2 a_2^2 +\sigma_3 a_3^2 &=& 0,
\\
\sigma_1 b_1^2 +\sigma_2 b_2^2 +\sigma_3 b_3^2 &=& 0,
\\
\sigma_1 c_1^2 +\sigma_2 c_2^2 +\sigma_3 c_3^2 &=& 0,
\eeqa
\end{subequations}
where $a_i ={\bd a} \cdot {\bd e}_i$, $i=1,2,3$,  are  components relative to the
orthonormal  principal axes of stress satisfying 
$[ {\bd e}_1, {\bd e}_2, {\bd e}_3]=1$, i.e. ${\bd a} = a_1{\bd e}_1 +a_2{\bd e}_2+a_3{\bd e}_3$. 
Equations \rf{131} imply  that each of the unit vectors lies on the curve $C$ defined by the intersection of an elliptical cone with the unit sphere: 
\beq{132}
C\equiv \big\{x_1{\bd e}_1 +x_2{\bd e}_2 + x_3{\bd e}_3\big\}:\quad 
\sigma_1 x_1^2 +\sigma_2 x_2^2 +\sigma_3 x_3^2 = 0,\quad x_1^2 +x_2^2 +x_3^2 =1.
\eeq
Alternatively, writing the three conditions as a matrix equation, 
\beq{02}
\begin{pmatrix}
a_1^2 & a_2^2 & a_3^2 
\\
b_1^2 & b_2^2 & b_3^2 
\\
c_1^2 & c_2^2 & c_3^2 
\end{pmatrix}
\begin{pmatrix}\sigma_1 \\ \sigma_2 \\ \sigma_3
\end{pmatrix}
=\begin{pmatrix}0 \\ 0 \\ 0 
\end{pmatrix}, 
\eeq
it is clear that  the following is a necessary condition for the triad $\{{\bd a},{\bd b},{\bd c}\}$ to simultaneously lie on $C$, 
\beq{133}
h( {\bd a},{\bd b},{\bd c}) \equiv 
\det
\begin{pmatrix}
a_1^2 & a_2^2 & a_3^2 
\\
b_1^2 & b_2^2 & b_3^2 
\\
c_1^2 & c_2^2 & c_3^2 
\end{pmatrix} = 0 . 
\eeq
We will show that if one member of the triad is known then this compatibility condition   defines the other two.   

The main result is as follows: 
\begin{thm}\label{t1}
The  triad $\{{\bd n}, {\bd v}_1({\bd n}),{\bd v}_2({\bd n})\}$  is orthonormal with $[{\bd n}, {\bd v}_1, {\bd v}_2]= 1$ and satisfies the compatibility condition for shear axes, i.e. $h({\bd n}, {\bd v}_1, {\bd v}_2)= 0$ (eq. \rf{133}). 
The vector functions ${\bd v}_1 $ and ${\bd v}_2$ are defined 
\beq{1aa}
{\bd v}_1 ({\bd n}) = \tfrac1{\sqrt{2}}\big( {\bd u}_- - s\,  {\bd u}_+),  
\qquad
{\bd v}_2 ({\bd n}) = \tfrac1{\sqrt{2}}\big( {\bd u}_- + s\,  {\bd u}_+),  
\eeq
where 
$$
s({\bd n}) = \sgn \big[ n_1n_2n_3 (n_1^2 - \tfrac13)(n_2^2 - \tfrac13)(n_3^2 - \tfrac13)\big],
$$
and  the orthogonal unit vectors ${\bd u}_\pm $  are   
\begin{subequations}\label{001}
\beqa{1b}
{\bd u}_\pm ({\bd n}) &=& 
\frac{\rho_\pm n_1}{n_2^2 - n_3^2 \pm  g  }\, {\bd e}_1 + 
\frac{\rho_\pm n_2}{n_3^2 - n_1^2 \pm  g  }\, {\bd e}_2 + 
\frac{\rho_\pm n_3}{n_1^2 - n_2^2 \pm  g  }\, {\bd e}_3, 
\\
\rho_\pm ({\bd n}) &=& \bigg[ 
\frac{n_1^2}{ \big( n_2^2 - n_3^2 \pm  g \big)^2 } + 
\frac{n_2^2}{ \big( n_3^2 - n_1^2 \pm  g \big)^2 } + 
\frac{n_3^2}{ \big( n_1^2 - n_2^2 \pm  g \big)^2}
\bigg]^{-1/2}, \label{1d}
\\
g({\bd n}) &=& \big[ 1 - 4(n_1^2n_2^2+n_2^2n_3^2+n_3^2n_1^2) +9n_1^2n_2^2n_3^2  \big]^{1/2}. 
\label{1c}
\eeqa
\end{subequations}
\end{thm}

The vector functions ${\bd v}_1$ and ${\bd v}_2$ together with $\bd{n}$ form an orthonormal triad, and therefore satisfy  the cyclic identities
\beq{435}
{\bd v}_1 ( {\bd n})  = {\bd v}_2 ( {\bd n}) \times {\bd n} ,
\qquad
{\bd v}_2 ( {\bd n}) = 
{\bd n} \times{\bd v}_1 ( {\bd n})    , 
\qquad
 {\bd n} = {\bd v}_1 ( {\bd n})\times {\bd v}_2 ( {\bd n}).
\eeq

An immediate corollary of Theorem 1 is that the triad $\{{\bd n},{\bd m},{\bd p}\}$ 
form a pure shear basis  if  one  lies on $C$ and the other two are  given by Theorem \ref{t1}.  For instance, if ${\bd n}$ satisfies 
\beq{03}
\sigma_1 n_1^2 +\sigma_2 n_2^2 +\sigma_3 n_3^2 = 0 ,  
\eeq
and ${\bd m}$ and ${\bd p} $ are defined by the vector functions,  
\beq{453}
{\bd m}={\bd v}_1 ({\bd n}),
\qquad 
{\bd p}={\bd v}_2 ({\bd n}),
\eeq
then   $\sigma_{nn}=\sigma_{mm}=\sigma_{pp}=0$.

Proof of Theorem \ref{t1}: Consider eq. \rf{133} for an orthonormal triad  $\{{\bd n},{\bd m},{\bd p}\}$.      
  Since the vectors are orthonormal   it follows that the sum of the three elements in each column and  each row of the determinant is unity. Thus, 
\beq{13}
f({\bd m})\equiv 
\left|
\begin{matrix}
n_1^2 & n_2^2 & n_3^2
\\
m_1^2 & m_2^2 & m_3^2
\\
1 & 1 & 1\end{matrix}\right| =
(n_3^2 - n_2^2)m_1^2 +
(n_1^2 - n_3^2)m_2^2 +
(n_2^2 - n_1^2)m_3^2 = 0,
\eeq
 is a  necessary but not sufficient condition for the transformation  from principal axes to pure shear axes.
We will now use this condition to find $\bd m$ and $\bd p$  for a given $\bd n$.
Consider the generalized function 
\beq{14}
F ({\bd u}) = f ({\bd u})-
  \lambda \,{\bd u}\cdot {\bd u}  + 2\rho\, {\bd n}\cdot {\bd u} .
\eeq
The form of $F$ is motivated by the compatibility condition for pure shear axes  augmented by terms that constrain the vector $\bd u$ to be of unit magnitude and orthogonal to $\bd n$.  The  Lagrange multipliers $\lambda$ and $\rho $ are determined by requiring that $f$ is stationary with respect to  ${\bd u}$ for a given ${\bd n}$.   Setting the gradient to zero  implies
\beq{6}
{\bd u} = 
\frac{\rho n_1}{n_2^2 - n_3^2 + \lambda}\, {\bd e}_1 + 
\frac{\rho n_2}{n_3^2 - n_1^2 + \lambda}\, {\bd e}_2 + 
\frac{\rho n_3}{n_1^2 - n_2^2 + \lambda}\, {\bd e}_3, 
\eeq
where $\rho$ is such that ${\bd u}$ is of unit magnitude, and $\lambda$ follows from the condition ${\bd n}\cdot {\bd u} = 0$, 
\beq{7}
\frac{ n_1^2}{n_2^2 - n_3^2 + \lambda} +
\frac{ n_2^2}{n_3^2 - n_1^2 + \lambda} +
\frac{ n_3^2}{n_1^2 - n_2^2 + \lambda} 
= 0 .
\eeq
This is a quadratic equation in $\lambda$ which can be simplified  using  $|{\bd n}|=1$  to the form 
\beq{1}
\lambda^2 =  1 - 4(n_1^2n_2^2+n_2^2n_3^2+n_3^2n_1^2) +9n_1^2n_2^2n_3^2 .  
\eeq
Thus $\lambda = \pm g({\bd n})$, where $g$ is defined in eq. \rf{1c}.   

Equations \rf{14} through \rf{7} imply that the function  $F({\bd u})$ vanishes at the stationary value of ${\bd u}$, and   hence 
\beq{15}
f ({\bd u}) =  \lambda .
\eeq 
The function  $f({\bd u})$ therefore achieves its maximum and minimum values $\pm g ({\bd n})$ along the directions  vectors ${\bd u}_\pm$ defined by \rf{1b}. It may also be checked, again using \rf{7}, that  ${\bd u}_+$ and ${\bd u}_-$ together with ${\bd n}$ form an orthonormal triad.  Furthermore, since $f ({\bd m}) $ is a quadratic form in the vector $\bd m$, it must have the form 
\beqa{16}
f ({\bd m}) &=& g ({\bd n})\big[ ({\bd m}\cdot{\bd u}_+)^2 - ({\bd m}\cdot{\bd u}_-)^2\big]
\nonumber 
\\
&=& -2g ({\bd n})\, {\bd m}\cdot{\bd v}_1({\bd n})\, {\bd m}\cdot{\bd v}_2({\bd n})
\qquad {\rm for }\,\,  {\bd m}\cdot {\bd n} =0, 
\eeqa
where  ${\bd v}_1$ and ${\bd v}_2$ are the orthogonal bisectors of the directions ${\bd u}_+$ and ${\bd u}_-$, and therefore correspond to the directions for which $f$ vanishes.  This proves the Theorem. 
 
 The parameter $s=\pm 1$ is introduced to ensure that the scalar triple product 
 $[{\bd n}, {\bd v}_1, {\bd v}_2] = {\bd n}\cdot({\bd v}_1 \times {\bd v}_2)$ is $+1$. This guarantees that the transformation from principal axes to the pure shear basis is a proper orthogonal transformation.

\subsection{Discussion}

\subsubsection{Stresses dual to the principal stresses}
Before discussing the vector functions in Theorem 1, we note some alternative formulations.  We first introduce the stress parameter  
$\sigma_{ (n)}$ for the pure shear axis  $\bd{n}$.  
Thus, using the pure shear definition \rf{82}, 
 the  condition \rf{03} for the vanishing stress $\sigma_{nn}$ may be written  
\beq{04}
\frac{n_2^2- n_3^2}{\sigma_1} = \frac{n_3^2- n_1^2}{\sigma_2} = \frac{n_1^2- n_2^2}{\sigma_3}  \equiv \frac{1}{ \sigma_{ (n)}}, 
\eeq
which defines $\sigma_{ (n)}$.  
Equivalently, 
\beq{91}
\begin{pmatrix}
n_1^2\\
n_2^2\\
n_3^2
\end{pmatrix}
=
\begin{pmatrix}
\frac13\\
\frac13\\
\frac13
\end{pmatrix}
+\frac{1}{3\sigma_{ (n)}}
\begin{pmatrix}
\sigma_3- \sigma_2 \\
\sigma_1- \sigma_3\\
\sigma_2- \sigma_1
\end{pmatrix}.
\eeq
The stress parameter $\sigma_{ (n)}$ can be related to the total traction $\tau_n$, defined by  $\tau_n^2 = |\bd{\sigma}\bd{n}|^2$ or using $\sigma_{nn}=0$, 
\beq{92}
\tau_n^2 = \sigma_1^2n_1^2+ \sigma_2^2n_2^2+ \sigma_3^2n_3^2. 
\eeq
Thus,
\beq{93}
\frac{1}{\sigma_{ (n)}}
= \frac{3\tau_n^2 - (\sigma_1^2+\sigma_2^2+\sigma_3^2 )}
{(\sigma_1 - \sigma_2)(\sigma_2 - \sigma_3)(\sigma_3 - \sigma_1)}, 
\eeq
and conversely
\beq{94}
\tau_n^2 = \frac13( \sigma_1^2 + \sigma_2^2 + \sigma_3^2 ) 
+\frac{1}{3\sigma_{ (n)}}(\sigma_1 - \sigma_2)(\sigma_2 - \sigma_3)(\sigma_3 - \sigma_1). 
\eeq
Note that elimination of $\sigma_{ (n)}$ from eq. \rf{91} gives the known relations for pure shear \cite{Ting06a}
\begin{align}\label{ting}
n_1^2 & = \frac{\tau_n^2 + \sigma_2\sigma_3}
{(\sigma_1 - \sigma_2)(\sigma_1 - \sigma_3)},
\nonumber \\
n_2^2 & = \frac{\tau_n^2 + \sigma_3\sigma_1}
{(\sigma_2 - \sigma_3)(\sigma_2 - \sigma_1)},
 \\
n_3^2 & = \frac{\tau_n^2 + \sigma_1\sigma_1}
{(\sigma_3 - \sigma_1)(\sigma_3 - \sigma_2)}. 
\nonumber 
\end{align}

Let   $\{{\bd a},{\bd b},{\bd c}\}$ be a pure shear basis  satisfying eq. 
\rf{01}, and define $\sigma_{ (a)}$, $\sigma_{ (b)}$ and  $\sigma_{ (c)}$ according to eq. \rf{04}.  It follows from eq. \rf{93} that
\beq{95}
\frac{1}{\sigma_{ (a)}} +\frac{1}{\sigma_{ (b)}}+\frac{1}{\sigma_{ (c)}}
= 3\frac{( \tau_a^2 +\tau_b^2+\tau_c^2 - \sigma_1^2 -\sigma_2^2 -\sigma_3^2 ) }
{(\sigma_1 - \sigma_2)(\sigma_2 - \sigma_3)(\sigma_3 - \sigma_1)}. 
\eeq
However, 
$\tau_a^2 +\tau_b^2+\tau_c^2 = \tr \big( [ \bd{a}\otimes \bd{a}+
\bd{b}\otimes \bd{b} + \bd{c}\otimes \bd{c}]\bd{\sigma}^2\big)
= \tr \bd{\sigma}^2$, 
 since $\{{\bd a},{\bd b},{\bd c}\}$ is an orthonormal basis. 
We therefore deduce the identity:
\beq{122}
\framebox[4.in]{
{\large $  \frac{1}{\sigma_{ (a)}} +\frac{1}{\sigma_{ (b)}}+\frac{1}{\sigma_{ (c)}}
= 0 \qquad $} \mbox{for a pure shear basis.} }
\eeq 
This condition for a pure shear basis should be compared with the condition 
\rf{82} satisfied by the  principal stresses.   The stress parameters 
$\sigma_{ (a)}$, $\sigma_{ (b)}$ and  $\sigma_{ (c)}$ are in this sense dual to the three principal stresses, and the condition \rf{122} is the dual of the pure shear condition $\tr \bd{\sigma} = 0$. 

\subsubsection{An alternative form of the vector functions}

Returning to the proof of Theorem 1, specifically the expression \rf{6} for the vector  $\bd{u}$ at a stationary value of the functional, we note that terms such as 
$(n_2^2-n_3^2)$ may be replaced by $\sigma_1 /\sigma_{ (n)}$ using eq. \rf{04}.  As a result,  stationary values of the vector  $\bd{u}$ are of the form 
\beq{86}
{\bd u} = 
\frac{\rho n_1}{\sigma_1 + \lambda}\, {\bd e}_1 + 
\frac{\rho n_2}{\sigma_2 + \lambda}\, {\bd e}_2 + 
\frac{\rho n_3}{\sigma_3 + \lambda}\, {\bd e}_3, 
\eeq
where $\lambda$ is now determined by the necessary condition 
${\bd n}\cdot {\bd u} = 0$, 
\beq{87}
\frac{ n_1^2}{\sigma_1 + \lambda} +
\frac{ n_2^2}{\sigma_2 + \lambda} +
\frac{ n_3^2}{\sigma_3 + \lambda} 
= 0 .
\eeq 
This can be simplified using  the fact that $\sigma_{nn}=0$ to  
\beq{71}
\lambda^2 +\sigma_2\sigma_3n_1^2 +\sigma_3\sigma_1n_2^2 +\sigma_1\sigma_2 n_3^2=0. 
\eeq
Consequently, the intermediate vectors ${\bd u}_\pm$ in Theorem 1 can be expressed
\begin{subequations}\label{007}
\beqa{301}
{\bd u}_\pm ({\bd n}) &=& 
\frac{\rho_\pm n_1}{\sigma_1 \pm  d  }\, {\bd e}_1 + 
\frac{\rho_\pm n_2}{\sigma_2 \pm  d  }\, {\bd e}_2 + 
\frac{\rho_\pm n_3}{\sigma_3 \pm  d  }\, {\bd e}_3, 
\\
\rho_\pm ({\bd n}) &=& \bigg[ 
\frac{n_1^2}{ \big( \sigma_1 \pm  d \big)^2 } + 
\frac{n_2^2}{ \big( \sigma_2 \pm  d \big)^2 } + 
\frac{n_3^2}{ \big( \sigma_3 \pm  d \big)^2}
\bigg]^{-1/2}, \label{011d}
\\
d({\bd n}) &=& \big[ - (n_1^2\sigma_2\sigma_3 +n_2^2\sigma_3\sigma_1+n_3^2\sigma_1\sigma_2) \big]^{1/2}.
\eeqa
\end{subequations}
These formulae provide an alternative  representation which combines the vector coordinates of $\bd{n}$ with the principal stress values.  We are now in a position to discuss the vector functions ${\bd v}_1$ and ${\bd v}_2$.

\subsubsection{The vector functions} 

\bigskip
\begin{figure}[htbp]
				\begin{center}	
 	
\setlength{\unitlength}{.2in}
\begin{picture}(10,10)(0,0)
\drawline[AHnb=0,linewidth=.05](0,0)(8,0)(8,8.764)(0,8.764)(0,0)
\drawline[AHnb=0,linewidth=.05](0,8.764)(4,10.764)(12,10.764)(12,2)(8,0)
\drawline[AHnb=0,linewidth=.05](8,8.764)(12,10.764)
\drawline[AHnb=0,linewidth=.1](4,4.382)(8,8.764)(8,4.382)(4,4.382)
\put(2.8,4.3){\makebox(0,0){\bf{\large{001}}}}
\put(9.1,4.3){\makebox(0,0){\bf{\large{110}}}}
\put(9.1,8.5){\makebox(0,0){\bf{\large{111}}}}

\end{picture}
	\caption{\small All directions are covered by considering  $\bd n$ restricted to the irreducible $1/48 $th of the cube surface shown. The region is  defined by the  triangle on the cube surface with vertices in the 001, 110 and 111 directions.  }
		\label{fcube} \end{center}  
	\end{figure}
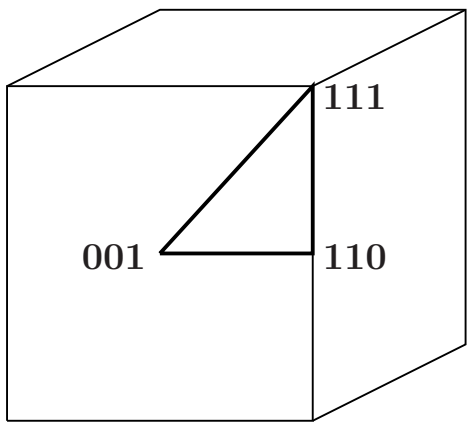

The vector functions ${\bd v}_1$ and ${\bd v}_2$ of Theorem 1 are expressed in two different forms through eqs. \rf{001} and \rf{007} 
in terms of the secondary vector functions ${\bd u}_\pm $.   It is useful to keep the  alternative representations in mind as we consider the dependence of the vector functions on $\bd n$ and on the principal stresses.  The range of variation of $\bd n$ may be reduced, with no loss in generality, to the interior of the triangle depicted in Figure \ref{fcube}.   The vertices $001$, $110$ and $111$,  described using  crystallographic notation, correspond to the generic directions $\bd{e}_i$,  $(\bd{e}_i+\bd{e}_j)/\sqrt{2}$ and $(\bd{e}_1+\bd{e}_2+\bd{e}_3)/\sqrt{3}$
($i\ne j=$1, 2 or 3).  

The functions ${\bd v}_1(\bd{n})$ and ${\bd v}_2(\bd{n})$ of Theorem 1 are well defined and unambiguous for $\bd{n}$ lying within the interior of the triangle although  some care is required when evaluating the functions on the boundary.  Consider, for instance the boundary between the 001 and 110 vertices, which is characterised by one of the rectangular components of $\bd n$ tending to zero.     
Consider   $n_3  = $o$(1)$ with  $(n_1^2 -n_2^2)=$O$(1)$, then to leading order in $n_3$ the vectors ${\bd u}_\pm $ 
are  $\bd{e}_3$ and $\bd{e}_3\times \bd{n}$.  This approximation holds for all $\bd{n}$ along the edge between 001 and 110 in Fig. \ref{fcube}, including    $\bd{n}$ approaching a principal axis (the 001 direction).  As $\bd{n}$ approaches a face diagonal (the 110 direction) the parameter  $g\rightarrow 0$ and consequently the function $f$ of eq. \rf{16} is identically zero for all vectors $\bd u$ in the plane orthogonal to $\bd n$. In this case, the vectors ${\bd u}_\pm $ and hence 
${\bd v}_1$ and ${\bd v}_2$ are degenerate.  Note that both the 001 and the 110 directions are pure shear axes if and only if one of the principal stresses vanishes.  Thus, if $\bd{n} = \bd{e}_3$ is a pure shear axis then $\sigma_{nn} = \sigma_3$ is zero.  Similarly, if  $\bd{n} = (\bd{e}_1+\bd{e}_2)/\sqrt{2}$  is a pure shear axis then $ \sigma_1+ \sigma_2=0$, which implies $\sigma_3=0$ since $\tr \bd{\sigma}=0$.  
We note that $0\le g\le 1$, with   $g({\bd e}_i)=1$,  $g(\frac1{\sqrt{2}}{\bd e}_i+\frac1{\sqrt{2}}{\bd e}_j)=0$ for $i\ne j$, and  $g(\frac1{\sqrt{3}}{\bd e}_1+\frac1{\sqrt{3}}{\bd e}_2+\frac1{\sqrt{3}}{\bd e}_3)=  0$.  In general, the function $g$ is zero only along face and cube diagonals. 

The 111 direction is a unique and special direction in terms of pure shear axes.  First, we note that regardless of the principal stresses,  111  is always a pure shear axis  by virtue of the constraint \rf{82}.  At the same time, the function $g(\bd{n})$ vanishes for $\bd{n}$ in the 111 direction, and the formula \rf{001} is not defined for this particular $\bd{n}$.  However, the alternative formulation of eq. \rf{007} yields 
\beq{308}
{\bd u}_\pm ({111}) = \frac{
\frac{{\bd e}_1 }{\sigma_1 \pm  \sigma_0  }  + 
\frac{{\bd e}_2  }{\sigma_2 \pm  \sigma_0  }  + 
\frac{{\bd e}_3  }{\sigma_3 \pm  \sigma_0  }  }
{\sqrt{ 
  \frac1{( \sigma_1 \pm  \sigma_0 )^2 } + 
  \frac1{( \sigma_2 \pm  \sigma_0 )^2 } + 
  \frac1{( \sigma_3 \pm  \sigma_0 )^2 }
}}, 
\eeq
where the positive stress $\sigma_0 \equiv d(111) = \big[ - ( \sigma_2\sigma_3 + \sigma_3\sigma_1+ \sigma_1\sigma_2)/3 \big]^{1/2}$ may be written  
\beq{348}
\sigma_0 = \frac1{\sqrt{2}} \tau_0 \qquad
\mbox{with } \qquad 
\tau_0  = \big[ \frac13 ( \sigma_1^2+\sigma_2^2+\sigma_3^2 ) \big]^{1/2}.
\eeq
$\tau_0$ 
is the total traction for the 111 direction ($\tau_0  = \tau_n$).
In summary, $({\bd u}_+ ({111}) \pm {\bd u}_- ({111}) )\sqrt{2}$ together with 111 define the pure shear basis.  

Using the general relation  
\beq{344}
g(\bd{n})= \frac{d(\bd{n})}{\sigma_{(n)} },
\eeq
 it follows that $1/\sigma_{(n)}$ vanishes in the 111 direction: 
 \beq{345}
\frac{1}{\sigma_{(111)} }= 0.
\eeq 
This is also apparent from the relation \rf{91} at $\bd{n}=111$.  Equation \rf{91}, which can be considered the definition of the dual stress $\sigma_{(n)}$, also provides a parameterization of possible directions $\bd{n}$ in terms of $\sigma_{(n)}$, considered as a variable.    The special direction 111 corresponds to 
$\sigma_{(n)}\rightarrow \infty$, and neighbouring $\bd{n}$ directions can be found by continuation.  By starting at the 111 direction one could develop all  pure shear bases for a given state of stress.  Thus, based on eq. \rf{92} let 
\beq{155}
\bd{n}(t) = \sqrt{\frac13 + \frac{t}{3} p_1}\, \bd{e}_1
+ \sqrt{\frac13 + \frac{t}{3} p_2 }\, \bd{e}_2+\sqrt{\frac13 + \frac{t}{3} p_3}\, \bd{e}_3, 
\eeq
where the unit vector $\bd{p}$ is defined by the principal stresses 
\beq{492}
\bd{p} = \frac{ (\sigma_3-\sigma_2)}{\tau_0} \bd{e}_1
+ \frac{(\sigma_1-\sigma_3)}{\tau_0} \bd{e}_2+\frac{ (\sigma_2-\sigma_1)}{\tau_0} \bd{e}_3. 
\eeq
The range of the parameter $t$ is such that the arguments of the square roots lie between $0$ and $1$.  The complete set of pure shear bases for a given state of stress is then characterised by the set of triads $\{ \bd{n}(t), \bd{v}_1(\bd{n}), \bd{v}_2(\bd{n})\}$ defined by all permissible $t$. 



\section{Strain energy in cubic materials}\label{sec3}

We now consider the relationship between the distinct types of basis for a state of pure shear and the  strain energy in a solid.  The strain energy $U$  is assumed to be  quadratic in the symmetric strain tensor
 ${\bd \varepsilon}$, but  using the linear relation between stress and strain the energy can be written as a quadratic in stress.  Our focus is   the dependence of the strain energy on the orientation of the principal  axes of the stress  ${\bd \sigma}$, which is no longer constrained to be a state of pure shear.  An alternative point of view is that  for a given state of stress or strain, we consider how the strain energy varies as the crystal axes are rotated.  The general question of how strain energy behaves in this manner and what are the optimal crystal orientations has been considered in several papers.  Rovati and  Taliercio \cite{Rovati03} provide a good review of the topic, and they also examine the particular case of materials with  cubic symmetry.  The author considered the same question \cite{Norris05}, and derived several results concerning   orientations of the cube axes that  yield  extremal values of strain energy.  The purpose of this section is to revisit this issue afresh, and show that the question of which orientations provide the extreme values of $U$ is easily and directly answered using the notion of pure shear bases.

The isotropic invariants of strain are $I_{\rm iso}( {\bd \varepsilon})  \equiv \tr {\bd \varepsilon}  $ and $II_{\rm iso}( {\bd \varepsilon}) \equiv \tr {\bd \varepsilon}^2  $, and invariants  with cubic symmetry are    $
I_{\rm cub}( {\bd \varepsilon}) \equiv \tr {\tens D}{\bd \varepsilon} = I_{\rm iso}( {\bd \varepsilon}) $ and  $II_{\rm cub}( {\bd \varepsilon}) \equiv \tr ({\tens D}{\bd \varepsilon})^2$.  Here, ${\tens D}$ is a fourth order tensor defined by the cube axes, which are assumed to be  $\{{\bd a},{\bd b},   {\bd c}\}$, 
\beq{40}
{\tens D} = {\bd a}\otimes{\bd a}\otimes{\bd a}\otimes{\bd a}
 +{\bd b}\otimes{\bd b}\otimes{\bd b}\otimes{\bd b}
 +{\bd c}\otimes{\bd c}\otimes{\bd c}\otimes{\bd c}\, . 
 \eeq
 The strain energy   in the linear elastic theory is   
\beq{41}
2U = \kappa I_{\rm iso}^2 ( {\bd \varepsilon}) + 2\mu_1 \big[ II_{\rm iso}({\bd \varepsilon})  - II_{\rm cub}({\bd \varepsilon}) \big] 
+ 2\mu_2 \big[ II_{\rm cub}({\bd \varepsilon})  - \frac13 I_{\rm iso}^2({\bd \varepsilon}) \big]  .  
\eeq
This is positive definite if and only if the three constants $\kappa$,  $\mu_1$ and $\mu_2$ are all positive, which is assumed. 
These parameters, called the ``principal elasticities" by Kelvin \cite{kelvin},  are related to  the  standard (Voigt)    notation for stiffness: $\kappa = (c_{11}+2c_{12})/3$,  $\mu_1 = c_{44}$, $\mu_2 = (c_{11}-c_{12})/2$.
The fourth order tensors of elastic stiffness and its inverse, compliance, follow from  above.  Thus, 
\beq{84}
{\bd \sigma} = {\tens C} {\bd \varepsilon}, \qquad {\bd \varepsilon}= {\tens S}{\bd \sigma}, 
\eeq
where ${\tens S}= {\tens C}^{-1}$ and 
\beq{85}
 {\tens C}^{\pm 1} = (3\kappa)^{\pm 1}  {\tens J} + (2\mu_1)^{\pm 1}  {\tens L}+ (2\mu_2)^{\pm 1}  {\tens M}. 
\eeq
The tensors are as follows:  ${\tens J}{\bd S} = \frac13 (\tr {\bd S} ){\bd I}$ for all ${\bd S}\in \Sym$, 
\beq{305}
{\tens L} = {\tens I} -{\tens D}, \qquad
{\tens M} = {\tens D} -{\tens J}, 
\eeq
and ${\tens I}$ is the fourth order identity, ${\tens I}{\bd S} =  {\bd S} $ for all ${\bd S}\in \Sym$, which can be written  
\beq{306}
{\tens I} = {\tens J} + {\tens L} +{\tens M}. 
\eeq
As a consequence, the  strain energy can be written in an alternative form as a quadratic function of the stress invariants:
\beq{42}
2U = \frac1{9\kappa} I_{\rm iso}^2 ( {\bd \sigma}) + \frac1{2\mu_1} \big[ II_{\rm iso}({\bd \sigma})  - II_{\rm cub}({\bd \sigma}) \big]
+ \frac1{2\mu_2} \big[ II_{\rm cub}({\bd \sigma})  - \frac13I_{\rm iso}^2({\bd \sigma}) \big]  .  
\eeq

The stress may be partitioned into hydrostatic and deviatoric parts, 
\beq{43}
{\bd \sigma} =  \frac13 I_{\rm iso} ( {\bd \sigma}) {\bd I} + {\bd \sigma}', 
\eeq
where the deviatoric stress satisfies $\tr {\bd \sigma}' = 0$, and hence can be considered as a pure shear. Using $I_{\rm cub} ( {\bd \sigma}) 
= I_{\rm iso} ( {\bd \sigma})$, it follows that 
\beq{44}
II_{\rm iso}({\bd \sigma})  - \frac13 I_{\rm iso}^2({\bd \sigma}) = II_{\rm iso}({\bd \sigma}' ),
\qquad 
II_{\rm cub}({\bd \sigma})  - \frac13 I_{\rm iso}^2({\bd \sigma}) = II_{\rm cub}({\bd \sigma}' ).
\eeq
Hence, 
\beq{45}
2U = \frac1{9\kappa} I_{\rm iso}^2 ( {\bd \sigma}) + \frac1{2\mu_1}  II_{\rm iso}({\bd \sigma}')  
+ \bigr(\frac1{2\mu_2} -\frac1{2\mu_1} \bigr) II_{\rm cub}({\bd \sigma}')  ,  
\eeq
and all the orientation dependence is through the final term. 
Note that 
\beq{46}
0\le 
II_{\rm cub}({\bd \sigma}')=
{\sigma_{aa} '}^2 + {\sigma_{bb} '}^2 + {\sigma_{cc} '}^2  \le  II_{\rm iso}({\bd \sigma}')
= {\sigma_1 '}^2 + {\sigma_2 '}^2 + {\sigma_3 '}^2 ,
\eeq
where $\sigma_{aa}' = {\bd a}\cdot ({\bd \sigma}' {\bd a})$  etc., and $\sigma_i '$ are the principal deviatoric stresses, satisfying 
$\sigma_1 '+\sigma_2 '+\sigma_3'=0$. 
The maximum value of $II_{\rm cub}({\bd \sigma}')$ is obtained when the cube axes $\{ {\bd a}, {\bd b}, {\bd c}\}$ coincides with the principal axes $\{ {\bd e}_1, {\bd e}_2, {\bd e}_3 \}$.   The minimum value $II_{\rm cub}({\bd \sigma}')= 0$ occurs when the cube axes $\{ {\bd a}, {\bd b}, {\bd c}\}$ form  a pure shear basis.  The extreme values of the strain energy are therefore
\beqa{47}
2U = \frac1{9\kappa} I_{\rm iso}^2 ( {\bd \sigma}) +   II_{\rm iso}({\bd \sigma}')  
\times \left\{
\begin{array}{ll}
\frac{1}{2\mu_1}, & \quad \{ {\bd a}, {\bd b}, {\bd c}\} = \mbox{pure shear axes},\\
& \\
\frac1{2\mu_2}, & \quad \{ {\bd a}, {\bd b}, {\bd c}\} = \mbox{principal axes}.
\end{array}
\right. 
\eeqa
These results are unchanged  by  transformations congruent with cubic symmetry, i.e. 
$\{ {\bd a}, {\bd b}, {\bd c}\}\rightarrow \{ -{\bd a}, {\bd b}, {\bd c}\}$, etc.  Hence we have the second main result: 
\begin{thm}\label{t2}
If $\mu_2> \mu_1  \, (\mu_2 <\mu_1)$ then the strain energy is maximum (minimum) when the cube axes 
$\{ {\bd a}, {\bd b}, {\bd c}\}$ are aligned with a pure shear basis for the deviatoric stress.  The strain energy is  minimum (maximum) when the cube axes 
 coincide with the principal axes of stress. 
\end{thm}
These results may be phrased in terms of strain rather than stress.    

The state of pure shear  may be  reconsidered in terms of  the cubic fourth order tensors.  Thus, a state of  pure shear,  $\tr {\bd \sigma} = 0$,  is equivalent to 
 \beq{501}
 {\tens J}{\bd \sigma} =0,\quad  
 {\tens K}{\bd \sigma} ={\bd \sigma} , 
 \qquad
 \Leftrightarrow \qquad \mbox{pure shear},
 \eeq
 where the fourth order tensor 
 ${\tens K}$ is 
 \beq{506}
 {\tens K} = {\tens I}-{\tens J}= {\tens L}+{\tens M}.
 \eeq
 We note in passing that a state of pure hydrostatic stress (pressure) is the dual of 
 eq. \rf{501}:  
 ${\tens J}{\bd \sigma} ={\bd \sigma}$, $ {\tens K}{\bd \sigma} =0 $.  Concentrating on pure shear states of stress, the second partition of eq. \rf{506} enables us to define the distinguished basis vectors as follows 
\begin{subequations}\label{88}
\begin{align}
&
{\tens L}{\bd \sigma} ={\bd \sigma},\quad  
 {\tens M}{\bd \sigma} =0 , 
 \qquad
 \Leftrightarrow \qquad \{ {\bd a}, {\bd b}, {\bd c}\} = \mbox{pure shear axes},\\
 & 
 {\tens L}{\bd \sigma} =0,\quad  
 {\tens M}{\bd \sigma} ={\bd \sigma} , 
 \qquad
 \Leftrightarrow \qquad \{ {\bd a}, {\bd b}, {\bd c}\} = \mbox{principal axes}.
\end{align}
\end{subequations}
The fourth order tensors ${\tens J}$, ${\tens K}$, ${\tens L}$ and ${\tens M}$ act on elements of Sym, which is a 6-dimensional space.  In this way, 
 $\dim {\tens J}=1$,
$\dim {\tens L}=3$, $\dim {\tens M}=2$ and
 $\dim {\tens K}=\dim {\tens L}+\dim {\tens L}=5$.  Thus, a state of pure shear, 
 \rf{501}, is characterized by 5 degrees of freedom, while hydrostatic stress has only a single degree of freedom.  For the special and complementary pure shear configurations of \rf{88}, the principal axes state is defined by two parameters, since $\sigma_1$,  $\sigma_2$ and $\sigma_3$ are not independent but satisfy the pure shear constraint  eq. \rf{82}.  The pure shear axes have an extra degree of freedom, which is reflected in the non-unique nature of  pure shear basis, and in the fact that  they define a one parameter family of coordinate systems, as discussed by B\v{e}lik and  Fosdick \cite{Belik98}.  
 
 Finally, the  relationship between the deviatoric stress and strain is simple at the energy extrema.  Thus, eqs. \rf{85} and \rf{88} imply that ${\bd \sigma}' = 2\mu_1{\bd \varepsilon}'$ if the cube axes are pure shear axes, and ${\bd \sigma}' = 2\mu_2{\bd \varepsilon}'$ when they are the principal axes of  stress and/or strain.  

\section*{Acknowledgment}  The  constructive comments of a referee are appreciated. 
 

\end{document}